# Observation of Dislocation Non-Hermitian Skin Effect

Wenquan Wu,[1†] Qicheng Zhang,[1†] Liangjun Qi,[1] Kun Zhang,[1] Shuaishuai Tong[1] and Chunyin Qiu[1,2*]

[1]*Key Laboratory of Artificial Micro- and Nano-Structures of Ministry of Education and School of Physics and Technology, Wuhan University, Wuhan 430072, China*
[2]*Wuhan Institute of Quantum Technology, Wuhan 430206, China*

[†]*These authors contributed equally: Wenquan Wu, Qicheng Zhang*
[*]*To whom correspondence should be addressed: cyqiu@whu.edu.cn*

**Abstract:** The non-Hermitian skin effect (NHSE), a striking phenomenon where a large number of states accumulate toward open boundaries, has garnered significant attention in both fundamental physics and emerging applications. Recent theoretical studies unveiled a distinctive dislocation NHSE by disentangling it from the established boundary NHSE, thereby bridging the gap between topological defects and non-Hermitian effects. In this Letter, we report the first experimental observation of the dislocation NHSE, achieved using an ingeniously designed nonreciprocal, torus-like acoustic lattice with two dislocations of opposite Burgers vectors. Our results show that the sound energy density inside the sample dramatically accumulates at one dislocation, while being unusually depleted at the other, a response distinct from all existing NHSE phenomena. This novel non-Hermitian effect not only probes the interplay between nontrivial defects and point-gap topology, but also holds promise for practical applications, such as the design of topological sound vacuum pumps.

*Introduction.* As two guiding principles in Hermitian topological physics, the traditional bulk-boundary correspondence [1-3] and bulk-defect correspondence [4-6] reveal a deep connection between nontrivial bulk topology and robustly localized states. However, these principles are challenged in non-Hermitian systems due to the extreme sensitivity of eigenenergy spectra and eigenstates to boundary conditions. Remarkably, the bulk-boundary correspondence has been reconceptualized through the groundbreaking discovery of non-Hermitian skin effect (NHSE), where a macroscopic number of bulk states localize toward the open boundaries [7-23]. More recently, the bulk-defect correspondence has been extended to non-Hermitian systems. This leads to a new class of defect-induced NHSEs, which significantly broaden the scope and variety of existing NHSE phenomena [24-30]. Among these, the dislocation NHSE, exhibiting the capability of topological defects to trap $\mathcal{O}(l)$ modes in two-dimensional (2D) non-Hermitian systems, is particularly intriguing, as it offers a novel probe for the pristine non-Hermitian topology [26-29].

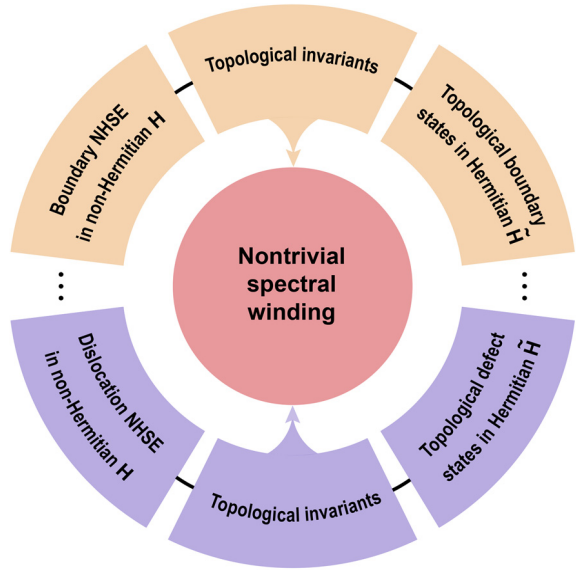

FIG. 1. Topological origins of the boundary NHSE and dislocation NHSE. Both phenomena can be attributed to the nontrivial point-gap topology, given the connection with the topological invariants of their Hermitian counterparts. Here $\mathbf{H}$ is a non-Hermitian Hamiltonian and $\widetilde{\mathbf{H}}$ is its parent Hermitian Hamiltonian constructed via a doubling procedure.

Physically, both the boundary NHSE and dislocation NHSE arise from the unique point-gap (or spectral winding) topology inherent in non-Hermitian systems. As illustrated in Fig. 1, for a given non-Hermitian Hamiltonian $\mathbf{H}$, one can always construct a parent Hermitian Hamiltonian $\widetilde{\mathbf{H}}$ with chiral symmetry through a doubling procedure [15]. Since these two Hamiltonians share the same algebraic topology, the topological invariant governing the nontrivial mid-gap edge states in $\widetilde{\mathbf{H}}$ is also responsible for the boundary NHSE in **H**, such as the winding number of the bulk spectrum in one-dimensional (1D) systems [15,16]. Similarly, the dislocation-induced NHSE in **H** and the topological defect states in $\widetilde{\mathbf{H}}$ can also be linked through the quantized indices defined based on nontrivial spectral windings and Burgers vectors of the real-space topological defects [26,27]. Remarkably, despite both closely-related to the point-gap topology, the dislocation NHSE and boundary NHSE exhibit entirely distinct morphologies of eigenstates. To date, various forms of the conventional boundary NHSE—such as the 1D NHSE [31-40], higher-order skin effect [41-46], hybrid skin-topological effect [47-52], and geometry-dependent skin effect [53-57]—have been successfully demonstrated across diverse experimental platforms. This greatly advances the non-Hermitian topological physics and inspires a wide range of innovative applications [31-34,37,38]. Equally important, it is of particular value to experimentally identify the dislocation NHSE, which bridges the gap between non-Hermitian spectral winding topology and crystalline defects [24-30].

In this Letter, we present the first experimental observation of dislocation NHSE by effectively coordinating the interplay between non-reciprocity, topological defects, and high-dimensional periodic boundaries. Theoretically, we consider a 2D weak Hatano-Nelson (WHN) lattice with two dislocations of opposite Burgers vectors, which exhibits a unique dislocation NHSE under full periodic boundary condition (PBC). Experimentally, the defective WHN lattice with PBC is emulated by a solidified, toroidal acoustic lattice equipped with active feedback circuits, where the latter accomplish the challenging nonreciprocal couplings in acoustics. Using an acoustic pumping-probe technique, we observe both dislocation skin and anti-skin effects [27,29], with sound energy accumulating around one dislocation while depleting near the other—an effect in stark contrast to those seen in the boundary NHSE. These results not only deepen our understanding of the interaction between nontrivial defects and point-gap topology, but also inspire potential applications in wave and energy manipulations.

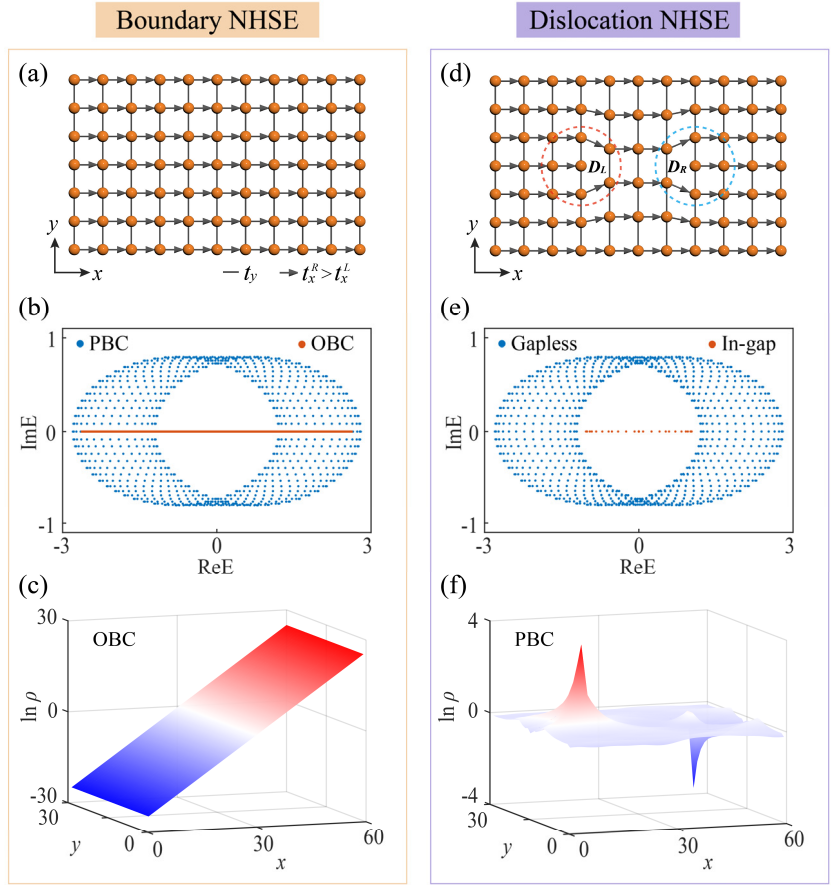

**Boundary NHSE**  **Dislocation NHSE**

FIG. 2. Dislocation NHSE and its comparison with boundary NHSE. (a) Schematic of a 2D WHN lattice, reciprocally coupled by *x*-directed 1D Hatano-Nelson chains in the *y* direction. (b) Complex energy spectra under PBC and OBC, calculated for a defect-free lattice of $2l \times l = 60 \times 30$ sites. (c) Logarithmic plot for the total probability density distribution of eigenstates under OBC, exhibiting a rightward, exponentially accumulated boundary NHSE. (d) Defective WHN lattice with a pair of dislocation cores, $D_L$ and $D_R$, formed by removing a row of $l = 30$ sites from the uniform lattice. (e) PBC spectrum of the defective lattice, including $\mathcal{O}(l^2)$ gapless extended states and $\mathcal{O}(l)$ in-gap localized states. (f) Dislocation NHSE manifested by the total probability density distribution of eigenstates. It exhibits a significant probability density accumulation (skin effect) and depletion (anti-skin effect) around the dislocation cores $D_L$ and $D_R$, respectively.

*Theoretical model.* As sketched in Fig. 2(a), we begin with a 2D weak WHN lattice that is simply stacked by 1D Hatano-Nelson chains [58]. It can be described by the momentum-space Bloch Hamiltonian

$$\mathbf{H}(k_x, k_y) = t_x^L e^{-ik_x} + t_x^R e^{ik_x} + 2t_y \cos k_y, \qquad (1)$$

where $t_y$ represents the reciprocal hopping in the *y* direction, while $t_x^L$ and $t_x^R$ correspond to the leftward and rightward hoppings in the *x* direction, respectively. Without loss of generality, we assume that the non-reciprocity $\lambda = t_x^R / t_x^L > 1$, as indicated by the rightward arrows in Fig. 2(a). Physically, the 2D WHN lattice can be viewed as a collection of 1D Hatano-Nelson chains with $k_y$-dependent real onsite energies $2t_y \cos k_y$. Therefore, its full PBC complex-energy spectrum forms a series of closed loops (each displaced along the real-energy axis), while the OBC spectrum consists of real eigenvalues within these loops. This is exemplified in Fig. 2(b) for a uniform lattice of $2l \times l = 60 \times 30$ sites, where the hopping parameters $t_x^L = 0.6$, $t_x^R = 1.4$, and $t_y = 0.3$. This nontrivial point-gap topology signals the presence of rightward NHSE at the boundary [15,16], as shown in Fig. 2(c). More quantitatively, the boundary NHSE can be characterized by an $x$-directed exponential accumulation of the total probability density $\rho(n_x, n_y) = \sum_m \left| \langle n_x, n_y | \psi_m^R \rangle \right|^2 = \lambda^{n_x - 1}$, where $| \psi_m^R \rangle$ represents the $m$th right eigenstate of the Hamiltonian, and $\langle n_x, n_y |$ is the position ket.

As sketched in Fig. 2(d), now we consider a 2D WHN lattice with a pair of dislocations $D_L$ and $D_R$, which are formed by removing a row of $l = 30$ sites using the conventional cut and glue procedure. To disentangle the system from boundary NHSE [26,27], we calculate its spectrum under full PBC in both $x$ and $y$ directions [Fig. 2(e)]. Compared with the PBC spectrum of 2D uniform WHN lattice [Fig. 2(b), blue dots], $\mathcal{O}(l)$ eigenenergies collapse onto the real energy axis [Fig. 2(e), red dots]. Originating from the nontrivial point-gap topology, these in-gap eigenstates are bound to the dislocations, whereas the remaining $\mathcal{O}(l^2)$ eigenstates appear to be extended across the lattice (see Supplemental Material). As displayed in Fig. 2(f), the total probability density of all eigenstates exhibits distinctive features of dislocation NHSE. It shows not only a rapid accumulation around $D_L$ but also a dramatic depletion near $D_R$, referred to as dislocation skin and anti-skin effects, respectively. In addition, although the vertical hopping $t_y$ is reciprocal, the $y$-directed probability density distribution is strongly nonuniform in the vicinity of dislocation cores. Both characteristics, which make the dislocation NHSE unique to higher-dimensional non-Hermitian defective lattices, can be intuitively explained as a result of non-equilibrium state pumping near the dislocation cores. That is, the number of arrows pointing into the core $D_L$ ($D_R$) exceeds (falls short of) those pointing out of it, leading to a net accumulation (depletion) of probability density at the corresponding core [27]. Note that the intriguing dislocation NHSE is topologically driven and robust in nature. It is protected by a nontrivial $\mathbb{Z}_2$ Hopf index

$$\vartheta = \hat{\mathbf{z}} \cdot (\mathbf{B} \times \boldsymbol{\nu})/2 \mod 1 = 1/2. \qquad (2)$$

Here $\mathbf{B} = (B_x, B_y) = (0, \pm 1)$ are Burgers vectors characterizing the topological defects at constant $y$, and the weak indices $\boldsymbol{\nu} = (1, 0)$ are 2D averaged winding numbers responsible for the rightward boundary NHSE in the $x$ direction. (In the doubled Hermitian Hamiltonian, $\vartheta = 1/2$ implies an odd parity of the topological dislocation zero modes.) The robustness is manifested in the fact that both the skin and anti-skin effects persist even as the reciprocal vertical coupling $t_y$ approaches zero.

By contrast, if changing the dislocation defects at constant $x$ [i.e., $\mathbf{B} = (\pm 1, 0)$], the index $\vartheta = 0$ suggests a trivial dislocation NHSE, where not only the anti-skin effect disappears, but also the robustness lost [27]. Without details presented here, we have examined the robustness of the dislocation NHSE against various types of disorder (see Supplemental Material).

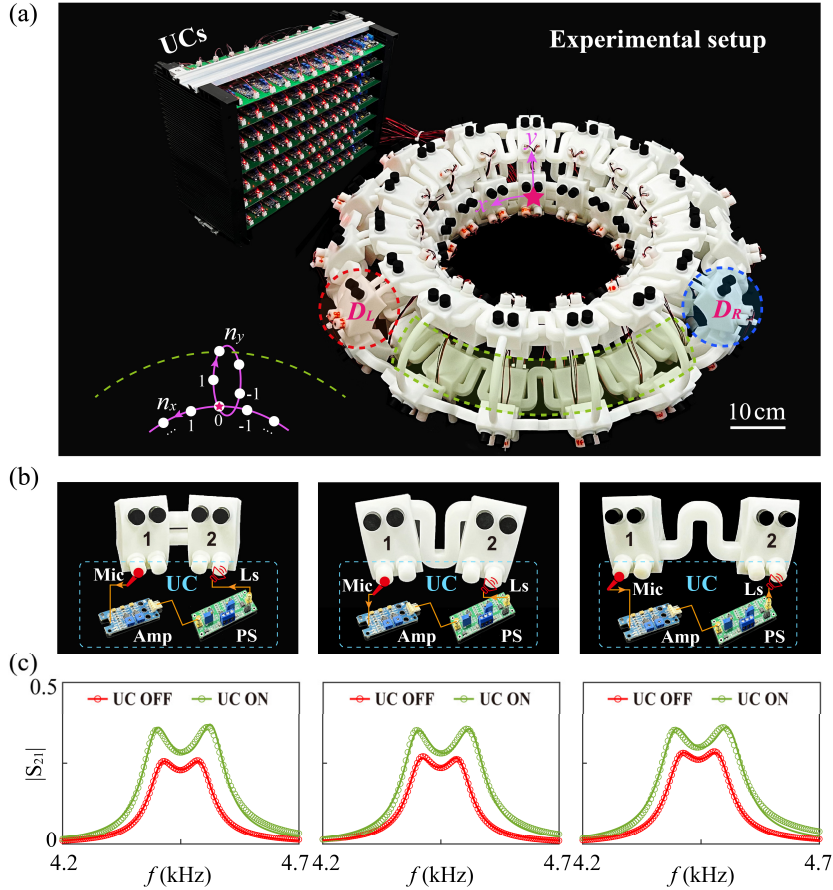

FIG. 3. Acoustic realization of the defective WHN lattice with full PBC. (a) Photograph of the experimental setup. It consists of a passive cavity-tube structure that simulates a reciprocal acoustic lattice, and a set of active unidirectional couplers (UCs) that implement additional unidirectional couplings. A defect line, ending with two dislocations $D_L$ and $D_R$, is created by removing four air-filled cavities from the layer of $n_y = 2$. The pink star denotes the position of the sound source, and the cavities are numbered as in the inset. (b) Three $x$-directed binary cavity-tube structures with different geometries but identical reciprocal couplings. Nonreciprocal couplings can be further achieved in this direction by integrating the UC circuits, each consisting of a microphone (Mic), an amplifier (Amp), a phase shifter (PS), and a loudspeaker (Ls). (c) Measured (circles) and fitted (lines) transmission responses $|S_{21}|$ in cavity 2, with the UC circuits switched on and off, to a source located in cavity 1.

*Experimental implementation of the defective acoustic WHN lattice with PBC.* Acoustic lattices have proven to be an excellent experimental platform for investigating both topological and non-Hermitian physics [37,44,46,52,55,57,59,60]. As shown in Fig. 3(a), to experimentally demonstrate the dislocation NHSE, we design a torus-like acoustic WHN lattice of $15 \times 5$ sites, in which a row of four sites is removed to create two dislocations of opposite Burgers vectors, $D_L$ and $D_R$. In the solidified acoustic structure, each air-filled cavity simulates a lattice site of chemical potential $f_0 \approx 4449$ Hz and an intrinsic loss $\gamma_0 \approx 45$ Hz, while the narrow tubes between neighboring cavities generate *reciprocal* acoustic couplings $t_x \approx 45$ Hz and $t_y \approx 30$ Hz. Meanwhile, active feedback circuits are introduced to realize extra rightward acoustic couplings $\kappa_x \approx 40$ Hz [59,60]. This, ultimately, gives rise to nonreciprocal acoustic couplings $t_x^L = t_x \approx 45$ Hz and $t_x^R = t_x + \kappa_x \approx 85$ Hz for the defective WHN lattice with PBC in both directions. Note that to coordinate the topological defects with torus geometry, we design five distinct types of elementary cavity-tube structures: three coupled along the $x$ direction and two coupled along the $y$ direction (see Supplemental Material for geometric details). As examples, Fig. 3(b) present the three $x$-directed binary cavity-tube structures, along with the unidirectional circuits for realizing nonreciprocal couplings. (The workflow of the unidirectional coupler can be introduced as follows: the sound signal in site 1 is picked up by a microphone, then modulated by an amplifier and a phase shifter, and finally output into the site 2 through a loudspeaker [59,60].) All acoustic parameters are calibrated by fitting the measured transmission spectra in these binary cavity-tube structures. As exemplified in Fig. 3(c), the experimental data for the three $x$-directed elementary structures are well reproduced using the same set of acoustic parameters, with the active unidirectional couplers either switched off ($\kappa_x = 0$) or switched on ($\kappa_x \approx 40$ Hz). Similar fitting results for the two $y$-directed binary cavity-tube structures are provided in Supplemental Material.

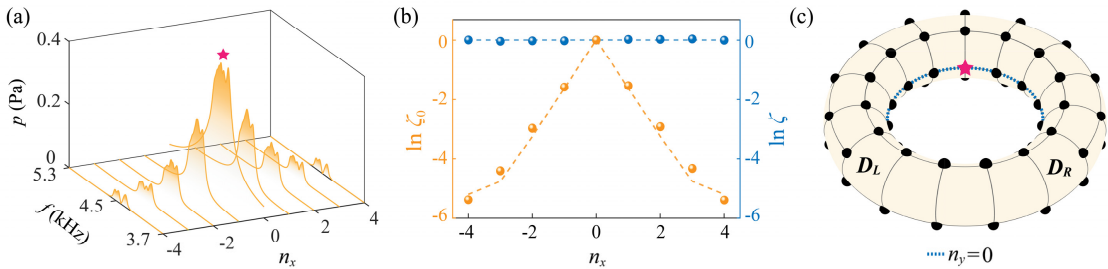

FIG. 4. Experimental examination of the uniform sound energy distribution in the reciprocal, torus-like acoustic lattice with two dislocations. In this case, all unidirectional circuits are turned off ($\kappa_x \approx 0$). (a) Spectral responses of sound pressure ($p$) measured for the cavities in the $n_y = 0$ layer. (b) Logarithmic plot of the sound energy density ($\ln\zeta_0$) distributed within this layer, along with the renormalized data ($\ln\zeta$) that counteract the influence of intrinsic loss. The experimental data (spheres)

reproduce well the numerical results based on coupled-mode theory (dashed lines). (c) Renormalized sound energy density $\zeta$, indicated by disk size, distributed across the sample.

*Experimental evidence for the dislocation NHSE.* Before demonstrating the dislocation-induced NHSE, we first examine the case without non-reciprocity ($\kappa_x = 0$). On the one hand, this provides an experimental basis and comparison for the subsequent nonreciprocal system; on the other hand, it allows us to assess the influence of intrinsic loss ($\gamma_0 \approx 45$ Hz), which often obscures the observation of NHSE [39,60]. To do this, we place a sound source at the position $(n_x, n_y) = (0,0)$ inside the cavity, and scan the sound responses across the sample. Figures 4(a) exemplifies the measured pressure spectra $p(f)$ for the layer $n_y = 0$, where the excitation frequency $f \in [3700, 5300]$ Hz. It shows that, although the eigenstates in this reciprocal system predict a uniform probability distribution, the undesired background loss results in a significant sound attenuation away from the source. To counteract this loss effect, we evaluate the sound energy $\zeta_0(n_x, n_y) = \sum_f |p(f)|^2$ for each cavity, and then renormalize the data as $\zeta(n_x, n_y) = \zeta_0(n_x, n_y)/\zeta_0(-n_x, n_y)$, since, intuitively, the symmetric cavities experience identical losses [60]. As shown in Fig. 4(b), the renormalized sound energy density (blue spheres) closely matches the expected uniform distribution (in the absence of non-reciprocity), in contrast to the approximately exponential decay observed in the original data (yellow spheres). Furthermore, Fig. 4(c) presents the renormalized sound energy density distributed across the entire sample, again exhibiting a nearly uniform distribution.

We now activate the feedback circuits to characterize the dislocation NHSE in our defective, toroidal acoustic WHN lattice and compare it with the conventional boundary NHSE in a uniform, open-boundary WHN lattice with identical hopping parameters. For easier comparison, the latter is also shaped into a torus but cut along specific rows and columns to create open boundaries (see Supplemental Material). As shown in Figs. 5(a) and 5(d), both the dislocations and open boundaries interact with the non-reciprocity of the systems, resulting in notable nonuniformities in the corresponding sound energy distributions. However, some significant differences can be observed between the dislocation NHSE and boundary NHSE. The most striking one is the dramatic energy accumulation and depletion occurring around the dislocation cores $D_L$ and $D_R$, which correspond to the unique skin and anti-skin effects in our defective WHN lattice, respectively. To demonstrate this more clearly, in Figs. 5(b) and 5(c) we plot the renormalized sound energy densities across several representative layers ($n_y = 0$ and $n_y = 2$) and columns ($n_x = \pm 5$). The experimental data show a rapid, non-exponential growth and decay of sound energy density near the topological dislocations, along with a relatively uniform sound distribution away from the defects (as the intact translational symmetry makes the dislocations nearly imperceptible). This forms a sharp contrast with the boundary NHSE in the defect-free WHN lattice under OBC: the sound energy density exhibits a standard exponential decay along the $x$-direction [Fig. 5(e)], while it remains rather uniform in the $y$-direction [Fig. 5(f)]. All these experimental data, which evidence the intriguing dislocation NHSE in our defective, torus-like acoustic WHN lattice, are consistent with the numerical calculations based on coupled-mode theory [61].

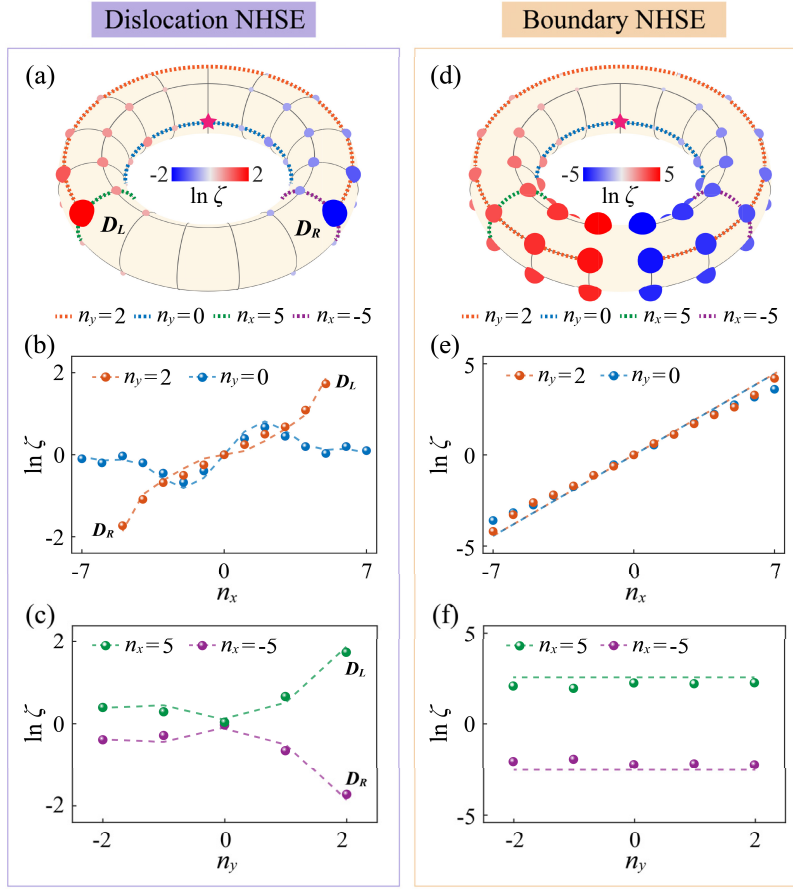

FIG. 5. Experimental evidence for the dislocation NHSE. (a) Renormalized sound energy density ($\zeta$) scanned across our defective, torus-like acoustic WHN lattice ($\kappa_x \approx$ 40 Hz), with the color bar indicating $\ln\zeta$ and disk size proportional to $|\ln\zeta|$. It clearly exhibits a sound accumulation around the dislocation $D_L$ and a depletion near the dislocation $D_R$, which correspond to a dislocation skin effect and a dislocation anti-skin effect, respectively. (b) Experimental data (spheres) extracted for the layers $n_y = 0$ and $n_y = 2$, matching well with the numerical results (dashed lines). (c) Experimental data extracted for the columns $n_x = \pm 5$. (d)-(f) Similar to (a)-(c), but for a uniform acoustic WHN lattice with open boundaries. The data exhibit a clear boundary NHSE: exponential in the $x$ direction while nearly uniform in the $y$-direction.

*Conclusions.* We have experimentally unveiled the highly intricate NHSE in a defective acoustic WHN lattice with PBC. In stark contrast to the extensively studied boundary NHSE, the dislocation NHSE exhibits distinctive observable features, including the rapid sound energy accumulation (skin effect) and depletion (anti-skin effect) near the dislocations, along with a relatively uniform sound energy distribution away from them. This phenomenon remarkably differs not only from the high-dimensional NHSE with edge or corner accumulations observed in previous works [44-46,50-52,55-57], but also from those defect localizations seen in Hermitian topological systems [62-73]. In particular, the topologically protected skin and anti-skin effects offer an unconventional mechanism for continuously pumping sound energy between dislocations, suggesting the possibility of a sound vacuum pump with exciting applications. Our findings expand the existing NHSE phenomena and pave the way for experimentally exploring a broad spectrum of non-Hermitian effects in conjunction with crystalline defects, such as disclination NHSE [25] and scale-free impurity localizations [74-76].

## Acknowledgements

This work was supported by the National Key R&D Program of China (Grant No. 2023YFA1406900), the National Natural Science Foundation of China (Grants No. 12104346, No. 12374418, and No. 12304495), the Natural Science Foundation of Hubei Province of China (No. 2024AFB654), and the Fundamental Research Funds for the Central Universities.